# The MADlib Analytics Library
*or MAD Skills, the SQL*

**Joseph M. Hellerstein**
hellerstein@berkeley.edu
U.C. Berkeley

**Christoper Ré**
chrisre@cs.wisc.edu
U. Wisconsin

**Florian Schoppmann**
Florian.Schoppmann@emc.com
Greenplum

**Daisy Zhe Wang**
daisyw@cise.ufl.edu
U. Florida

**Eugene Fratkin**
Eugene.Fratkin@emc.com
Greenplum

**Aleksander Gorajek**
Aleksander.Gorajek@emc.com
Greenplum

**Kee Siong Ng**
KeeSiong.Ng@emc.com
Greenplum

**Caleb Welton**
Caleb.Welton@emc.com
Greenplum

**Xixuan Feng**
xfeng@cs.wisc.edu
U. Wisconsin

**Kun Li**
kli@cise.ufl.edu
U. Florida

**Arun Kumar**
arun@cs.wisc.edu
U. Wisconsin

## ABSTRACT
MADlib is a free, open-source library of in-database analytic methods. It provides an evolving suite of SQL-based algorithms for machine learning, data mining and statistics that run at scale within a database engine, with no need for data import/export to other tools. The goal is for MADlib to eventually serve a role for scalable database systems that is similar to the CRAN library for R: a community repository of statistical methods, this time written with scale and parallelism in mind.

In this paper we introduce the MADlib project, including the background that led to its beginnings, and the motivation for its open-source nature. We provide an overview of the library's architecture and design patterns, and provide a description of various statistical methods in that context. We include performance and speedup results of a core design pattern from one of those methods over the Greenplum parallel DBMS on a modest-sized test cluster. We then report on two initial efforts at incorporating academic research into MADlib, which is one of the project's goals.

MADlib is freely available at http://madlib.net, and the project is open for contributions of both new methods, and ports to additional database platforms.

## 1. INTRODUCTION: FROM WAREHOUSING TO SCIENCE

Until fairly recently, large databases were used mainly for accounting purposes in enterprises, supporting financial record-keeping and reporting at various levels of granularity. *Data Warehousing* was the name given to industry practices for these database workloads. Accounting, by definition, involves significant care and attention to detail. Data Warehousing practices followed suit by encouraging careful and comprehensive database design, and by following exacting policies regarding the quality of data loaded into the database.

Attitudes toward large databases have changed quickly in the past decade, as the focus of usage has shifted from accountancy to analytics. The need for correct accounting and data warehousing practice has not gone away, but it is becoming a shrinking fraction of the volume—and the value—of large-scale data management. The emerging trend focuses on the use of a wide range of potentially noisy data to support predictive analytics, provided via statistical models and algorithms. *Data Science* is a name that is gaining currency for the industry practices evolving around these workloads.

Data scientists make use of database engines in a very different way than traditional data warehousing professionals. Rather than carefully designing global schemas and "repelling" data until it is integrated, they load data into private schemas in whatever form is convenient. Rather than focusing on simple OLAP-style drill-down reports, they implement rich statistical models and algorithms in the database, using extensible SQL as a language for orchestrating data movement between disk, memory, and multiple parallel machines. In short, for data scientists a DBMS is a scalable analytics runtime—one that is conveniently compatible with the database systems widely used for transactions and accounting.

In 2008, a group of us from the database industry, consultancy, academia, and end-user analytics got together to describe this usage pattern as we observed it in the field. We dubbed it **MAD**, an acronym for the *Magnetic* (as opposed to repellent) aspect of the platform, the *Agile* design patterns used for modeling, loading and iterating on data, and the *Deep* statistical models and algorithms being used. The "MAD Skills" paper that resulted described this pattern, and included a number of non-trivial analytics techniques implemented as simple SQL scripts [11].

After the publication of the paper, significant interest emerged not only in its design aspects, but also in the actual SQL implementations of statistical methods. This interest came from many directions: customers were requesting it of consultants and vendors, and academics were increasingly publishing papers on the topic. What was missing was a software framework to focus the energy of the community, and connect the various interested constituencies. This led to the design of **MADlib**, the subject of this paper.

### Introducing MADlib
MADlib is a library of analytic methods that can be installed and executed within a relational database engine that supports extensible SQL. A snapshot of the current contents of MADlib including methods and ports is provided in Table 1. This set of methods and ports is intended to grow over time.

| Category | Method |
|---|---|
| Supervised Learning | Linear Regression |
| | Logistic Regression |
| | Naive Bayes Classification |
| | Decision Trees (C4.5) |
| | Support Vector Machines |
| Unsupervised Learning | k-Means Clustering |
| | SVD Matrix Factorization |
| | Latent Dirichlet Allocation |
| | Association Rules |
| Decriptive Statistics | Count-Min Sketch |
| | Flajolet-Martin Sketch |
| | Data Profiling |
| | Quantiles |
| Support Modules | Sparse Vectors |
| | Array Operations |
| | Conjugate Gradient Optimization |

Table 1: Methods provided in MADlib v0.3. This version has been tested on two DBMS platforms: PostgreSQL (single-node open source) and Greenplum Database (massively parallel commercial system, free and fully-functional for research use.)

The methods in MADlib are designed both for in- or out-of-core execution, and for the shared-nothing, "scale-out" parallelism offered by modern parallel database engines, ensuring that computation is done close to the data. The core functionality is written in declarative SQL statements, which orchestrate data movement to and from disk, and across networked machines. Single-node inner loops use SQL extensibility to call out to high-performance math libraries in user-defined scalar and aggregate functions. At the highest level, tasks that require iteration and/or structure definition are coded in Python driver routines, which are used only to kick off the data-rich computations that happen within the database engine.

MADlib is hosted publicly at github, and readers are encouraged to browse the code and documentation via the MADlib website http://madlib.net. The initial MADlib codebase reflects contributions from both industry (Greenplum) and academia (UC Berkeley, the University of Wisconsin, and the University of Florida). Code management and Quality Assurance efforts have been contributed by Greenplum. At this time, the project is ready to consider contributions from additional parties, including both new methods and ports to new platforms.

## 2. GOALS OF THE PROJECT

The primary goal of the MADlib open-source project is to accelerate innovation and technology transfer in the Data Science community via a shared library of scalable in-database analytics, much as the CRAN library serves the R community [34]. Unlike CRAN, which is customized to the R analytics engine, we hope that MADlib's grounding in standard SQL can lead to community ports to a variety of parallel database engines.

### 2.1 Why Databases?

For decades, statistical packages like SAS, Matlab and R have been the key tools for deep analytics, and the practices surrounding these tools have been elevated into widely-used traditional methodologies. One standard analytics methodology advocated in this domain is called SEMMA: Sample, Explore, Modify, Model, Assess. The "EMMA" portion of this cycle identifies a set of fundamental tasks that an analyst needs to perform, but the first, "S" step makes less and less sense in many settings today. The costs of computation and storage are increasingly cheap, and entire data sets can often be processed efficiently by a cluster of computers. Meanwhile, competition for extracting value from data has become increasingly refined. Consider fiercely competitive application domains like online advertising or politics. It is of course important to target "typical" people (customers, voters) that would be captured by sampling the database. But the fact that SEMMA is standard practice means that optimizing for a sample provides no real competitive advantage. Winning today requires extracting advantages in the long tail of "special interests", a practice known as "microtargeting", "hypertargeting" or "narrowcasting". In that context, the first step of SEMMA essentially defeats the remaining four steps, leading to simplistic, generalized decision-making that may not translate well to small populations in the tail of the distribution. In the era of "Big Data", this argument for enhanced attention to long tails applies to an increasing range of use cases.

Driven in part by this observation, momentum has been gathering around efforts to develop scalable full-dataset analytics. One popular alternative is to push the statistical methods directly into new parallel processing platforms—notably, Apache Hadoop. For example, the open-source Mahout project aims to implement machine learning tools within Hadoop, harnessing interest in both academia and industry [10, 3]. This is certainly a plausible path to a solution, and Hadoop is being advocated as a promising approach even by major database players, including EMC/Greenplum, Oracle and Microsoft.

At the same time that the Hadoop ecosystem has been evolving, the SQL-based analytics ecosystem has grown rapidly as well, and large volumes of valuable data are likely to pour into SQL systems for many years to come. There is a rich ecosystem of tools, know-how, and organizational requirements that encourage this. For these cases, it would be helpful to push statistical methods into the DBMS. And as we will see, massively parallel databases form a surprisingly useful platform for sophisticated analytics. MADlib currently targets this environment of in-database analytics.

### 2.2 Why Open Source?

From the beginning, MADlib was designed as an open-source project with corporate backing, rather than a closed-source corporate effort with academic consulting. This decision was motivated by a number of factors, including the following:

- **The benefits of customization**: Statistical methods are rarely used as turnkey solutions. As a result, it is common for data scientists to want to modify and adapt canonical models and methods (e.g., regression, classification, clustering) to their own purposes. This is a very tangible benefit of open-source libraries over traditional closed-source packages. Moreover, in an open-source community there is a process and a set of positive incentives for useful modifications to be shared back to the benefit of the entire community.

- **Valuable data vs. valuable software**: In many emerging business sectors, the corporate value is captured in the data itself, not in the software used to analyze that data. Indeed, it is in the interest of these companies to have the open-source community adopt and improve their software. Open-source efforts can also be synergistic for vendors that sell commercial software, as evidenced by companies like EMC/Greenplum, Oracle, Microsoft and others beginning to provide Apache Hadoop alongside their commercial databases. Most IT shops today run a mix of open source and proprietary software, and many software vendors are finding it wise to position themselves intelligently in that context.

- **Closing the research-to-adoption loop**: Very few traditional database customers have the capacity to develop significant in-house research into computing or data science. On the other hand, it is hard for academics doing computing research to understand and influence the way that analytic processes are done in the field. An open-source project like MADlib has the potential to connect academic researchers not only to industrial software vendors, but also directly to the end-users of analytics software. This can improve technology transfer from academia into practice without requiring database software vendors to serve as middlemen. It can similarly enable end-users in specific application domains to influence the research agenda in academia.

- **Leveling the playing field, encouraging innovation**: Over the past two decades, database software vendors have developed proprietary data mining toolkits consisting of textbook algorithms. It is hard to assess their relative merits. Meanwhile, other communities in machine learning and internet advertising have also been busily innovating, but their code is typically not well packaged for reuse, and the code that is available was not written to run in a database system. None of these projects has demonstrated the vibrancy and breadth we see in the open-source community surrounding R and its CRAN package. The goal of MADlib is to fill this gap: bring the database community up to a baseline level of competence on standard statistical algorithms, remove opportunities for proprietary *FUD*, and help focus a large community on innovation and technology transfer.

## 2.3 A Model for Open-Source Collaboration

The design of MADlib comes at a time when the connections between open-source software and academic research seem particularly frayed. MADlib is designed in part as an experiment in binding these communities more tightly, to face current realities in software development.

In previous decades, many important open-source packages originated in universities and evolved into significant commercial products. Examples include the Ingres and Postgres database systems, the BSD UNIX and Mach operating systems, the X Window user interfaces and the Kerberos authentication suite. These projects were characterized by aggressive application of new research ideas, captured in workable but fairly raw public releases that matured slowly with the help of communities outside the university. While all of the above examples were incorporated into commercial products, many of those efforts emerged years or decades after the initial open-source releases, and often with significant changes.

Today, we expect successful open-source projects to be quite mature, often comparable to commercial products. To achieve this level of maturity, most successful open-source projects have one or more major corporate backers who pay some number of committers and provide professional support for Quality Assurance (QA). This kind of investment is typically made in familiar software packages, not academic research projects. Many of the most popular examples—Hadoop, Linux, OpenOffice—began as efforts to produce open-source alternatives to well-identified, pre-existing commercial efforts.

MADlib is making an explicit effort to explore a new model for industry support of academic research via open source. Many academic research projects are generously supported by financial grants and gifts from companies. In MADlib, the corporate donation has largely consisted of a commitment to allocate significant professional software engineering time to bootstrap an open-source sandbox for academic research and tech transfer to practice. This leverages a strength of industry that is not easily replicated by government and other non-profit funding sources. Companies can recruit high-quality, experienced software engineers with the attraction of well-compensated, long-term career paths. Equally important, software shops can offer an entire software engineering pipeline that cannot be replicated on campus: this includes QA processes and QA engineering staff. The hope is that the corporate staffing of research projects like MADlib can enable more impactful academic open-source research, and speed technology transfer to industry.

## 2.4 MADlib Status

MADlib is still young, currently (as of March, 2012) at Version 0.3. The initial versions have focused on establishing a baseline of useful functionality, while laying the groundwork for future evolution. Initial development began with the non-trivial work of building the general-purpose framework described in Section 3. Additionally, we wanted robust implementations of textbook methods that were most frequently requested from customers we met through contacts at Greenplum. Finally, we wanted to validate MADlib as a research vehicle, by fostering a small number of university groups working in the area to experiment with the platform and get their code disseminated (Section 5).

## 3. MADLIB ARCHITECTURE

The core of traditional SQL—`SELECT... FROM... WHERE... GROUP BY`—is quite a powerful harness for orchestrating bulk data processing across one or many processors and disks. It is also a portable, native language supported by a range of widely-deployed open-source and commercial database engines. This makes SQL an attractive framework for writing data-intensive programs. Ideally, we would like MADlib methods to be written entirely in straightforward and portable SQL. Unfortunately, the portable core of "vanilla" SQL is often not quite enough to express the kinds of algorithms needed for advanced analytics.

Many statistical methods boil down to linear-algebra expressions over matrices. For relational databases to operate over very large matrices, this presents challenges at two scales. At a macroscopic scale, the matrices must be intelligently partitioned into chunks that can fit in memory on a single node. Once partitioned, the pieces can be keyed in such a way that SQL constructs can be used to orchestrate the movement of these chunks into and out of memory across one or many machines. At a microscopic scale, the database engine must invoke efficient linear-algebra routines on the pieces of data it gets in core. To this end it has to have the ability to very quickly invoke well-tuned linear-algebra methods.

We proceed to discuss issues involved at both of these levels in a bit more detail, and solutions we chose to implement in MADlib.

### 3.1 Macro-Programming (Orchestration)

A scalable method for linear algebra depends upon divide-and-conquer techniques: intelligent partitioning of the matrix, and a pattern to process the pieces and merge results back together. This partitioning and dataflow is currently outside the scope of a traditional query optimizer or database design tool. But there is a rich literature from scientific computing on these issues (e.g., [9]) that database programmers can use to craft efficient in-database implementations. Once data is properly partitioned, database engines shine at orchestrating the resulting data movement of partitions and the piecewise results of computation.

#### 3.1.1 User-Defined Aggregation

The most basic building block in the macro-programming of MADlib is the use of user-defined aggregates (UDAs). In general,

aggregates—and the related window functions—are the natural way in SQL to implement mathematical functions that take as input the values of an arbitrary number of rows (tuples). DBMSs typically implement aggregates as data-parallel streaming algorithms. And there is a large body of recent work on online learning algorithms and model-averaging techniques that fit the computational model of aggregates well (see, e.g., [47]).

Unfortunately, extension interfaces for user-defined aggregates vary widely across vendors and open-source systems. Nonetheless, the aggregation paradigm (or in functional programming terms, "fold" or "reduce") is natural and ubiquitous, and we expect the basic algorithmic patterns for user-defined aggregates to be very portable. In most widely-used DBMSs (e.g., in PostgreSQL, MySQL, Greenplum, Oracle, SQL Server, Teradata), a user-defined aggregate consists of a well-known pattern of two or three user-defined functions:

1. A *transition function* that takes the current transition state and a new data point. It combines both into into a new transition state.
2. An optional *merge function* that takes two transition states and computes a new combined transition state. This function is only needed for parallel execution.
3. A *final function* that takes a transition state and transforms it into the output value.

Clearly, a user-defined aggregate is inherently data-parallel if the transition function is associative and the merge function returns the same result as if the transition function was called repeatedly for every individual element in the second state.

Unfortunately, user-defined aggregates are not enough. In designing the high-level orchestration of data movement for analytics, we ran across two main limitations in standard SQL that we describe next. We addressed both these limitations using driver code written in simple script-based user-defined functions (UDFs), which in turn kick off more involved SQL queries. When implemented correctly, the performance of the scripting language code is not critical, since its logic is invoked only occasionally to kick off much larger bulk tasks that are executed by the core database engine.

### 3.1.2 Driver Functions for Multipass Iteration

The first problem we faced is the prevalence of "iterative" algorithms for many methods in statistics, which make many passes over a data set. Common examples include optimization methods like Gradient Descent and Markov Chain Monte Carlo (MCMC) simulation in which the number of iterations is determined by a data-dependent stopping condition at the end of each round. There are multiple SQL-based workarounds for this problem, whose applicability depends on the context.

**Counted Iteration via Virtual Tables**. In order to drive a fixed number $n$ of independent iterations, it is often simplest (and very efficient) to declare a virtual table with $n$ rows (e.g., via PostgreSQL's `generate_series` table function), and join it with a view representing a single iteration. This approach was used to implement $m$-of-$n$ Bootstrap sampling in the original MAD Skills paper [11]. It is supported in some fashion in a variety of DBMSs, sometimes by writing a simple table function consisting of a few lines of code.
**Window Aggregates for Stateful Iteration**. For settings where the current iteration depends on previous iterations, SQL's windowed aggregate feature can be used to carry state across iterations. Wang et al. took this approach to implement in-database MCMC inference [43] (Section 5.2). Unfortunately the level of support for window aggregates varies across SQL engines.
**Recursive Queries**. Most generally, it is possible to use the recursion features of SQL to perform iteration with arbitrary stopping conditions—this was used by Wang et al. to implement Viterbi inference [44] (Section 5.2). Unfortunately, like windowed aggregates, recursion support in SQL varies across database products, and does not form a reliable basis for portability.
**Driver Functions**. None of the above methods provides both generality and portability. As a result, in MADlib we chose to implement complex iterative methods by writing a driver UDF in Python to control iteration, which passes state across iterations intelligently. A standard pitfall in this style of programming is for the driver code to pull a large amount of data out of the database; this becomes a scalability bottleneck since the driver code typically does not parallelize and hence pulls all data to a single node. We avoid this via a design pattern in which the driver UDF kicks off each iteration and stages any inter-iteration output into a temporary table via `CREATE TEMP TABLE... AS SELECT...` It then reuses the resulting temp table in subsequent iterations as needed. Final outputs are also often stored in temp tables unless they are small, and can be interrogated using small aggregate queries as needed. As a result, all large-data movement is done within the database engine and its buffer pool. Database engines typically provide efficient parallelism as well as buffering and spill files on disk for large temp tables, so this pattern is quite efficient in practice. Sections 4.2 and 4.3 provide discussion of this pattern in the context of specific algorithms.

### 3.1.3 Templated Queries

A second problem is a limitation of SQL's roots in first-order logic, which requires that queries be cognizant of the schema of their input tables, and produce output tables with a fixed schema. In many cases we want to write "templated" queries that work over arbitrary schemas, with the details of arity, column names and types to be filled in later.

For example, the MADlib `profile` module takes an arbitrary table as input, producing univariate summary statistics for each of its columns. The input schema to this module is not fixed, and the output schema is a function of the input schema (a certain number of output columns for each input column). To address this issue, we use Python UDFs to interrogate the database catalog for details of input tables, and then synthesize customized SQL queries based on templates to produce outputs. Simpler versions of this issue arise in most of our iterative algorithms.

Unfortunately, templated SQL relies on identifiers or expressions passed as strings to represent database objects like tables. As such, the DBMS backend will discover syntactical errors only when the generated SQL is executed, often leading to error messages that are enigmatic to the user. As a result, templated SQL necessitates that MADlib code perform additional validation and error handling up front, in order to not compromise usability. In the future we plan to support this pattern as a Python library that ships with MADlib and provides useful programmer APIs and user feedback.

## 3.2 Micro-Programming: Data Representations and Inner Loops

In addition to doing the coarse-grained orchestration of chunks, the database engine must very efficiently invoke the single-node code that performs arithmetic on those chunks. For UDFs that operate at the row level (perhaps called multiple times per row), the standard practice is to implement them in C or C++. When computing dense matrix operations, these functions would make native calls to an open-source library like LAPACK [2] or Eigen [17].

Sparse matrices are not as well-handled by standard math libraries, and require more customization for efficient representations both on disk and in memory. We chose to write our own sparse matrix library in C for MADlib, which implements a run-length encoding scheme. Both of these solutions require careful low-level coding, and formed part of the overhead of getting MADlib started.

The specifics of a given method's linear algebra can be coded in a low-level way using loops of basic arithmetic in a language like C, but it is nicer if they can be expressed in a higher-level syntax that captures the semantics of the linear algebra at the level of matrices and arrays. We turn to this issue next.

### 3.3 A C++ Abstraction Layer for UDFs

There are a number of complexities involved in writing C or C++-based user-defined functions over a legacy DBMS like PostgreSQL, all of which can get in the way of maintainable, portable application logic. This complexity can be especially frustrating for routines whose pseudocode amounts to a short linear-algebra expression that *should* result in a compact implementation. MADlib provides a C++ abstraction layer both to ease the burden of writing high-performance UDFs, and to encapsulate DBMS-specific logic inside the abstraction layer, rather than spreading the cost of porting across all the UDFs in the library. In brief, the MADlib C++ abstraction provides three classes of functionality: type bridging, resource management shims, and math library integration.

Type bridging is provided via an encapsulated mapping of C++ types and methods to database types and functions. UDFs can be written with standard C++ atomic types, as well as the vector and matrix types that are native to a high-performance linear-algebra library. We have successfully layered multiple alternative libraries under this interface, and are currently using Eigen [17] (see Section 4.4 for performance results and lessons learned about integrating linear-algebra libraries in a RDBMS). The translation to and from database types (including composite or array types like `double precision[]` for vectors) is handled by the abstraction layer. Similarly, higher-order functions in C++ can be mapped to the appropriate object IDs of UDFs in the database, with the abstraction layer taking care of looking up the function in the database catalog, verifying argument lists, ensuring type-safety, etc.

The second aspect of the C++ abstraction layer is to provide a safe and robust standard runtime interface to DBMS-managed resources. This includes layering C++ object allocation/deallocation over DBMS-managed memory interfaces, providing shims between C++ exception handling and DBMS handlers, and correctly propagating system signals to and from the DBMS.

Finally, by incorporating proven third-party libraries, the C++ abstraction layer makes it easy for MADlib developers to write correct and performant code. For example, the Eigen linear-algebra library contains well-tested and well-tuned code that makes use of the SIMD instruction sets (like SSE) found in today's CPUs. Likewise, the abstraction layer itself has been tuned for efficient value marshalling, and code based on it will automatically benefit from future improvements. By virtue of being a template library, the runtime and abstraction overhead is reduced to a minimum.

As an illustration of the high-level code one can write over our abstraction layer, Listings 1 and 2 show reduced, but fully functional code snippets that implement multiple linear regression (as discussed further in Section 4.1).

## 4. EXAMPLES

To illustrate the above points, we look at three different algorithmic scenarios. The first is Linear Regression using Ordinary Least Squares (OLS), which is an example of a widely useful, simple single-pass aggregation technique. The second is binary Logistic Regression, another widely used technique, but one that employs an iterative algorithm. Finally, we look at $k$-Means Clustering, an iterative algorithm with large intermediate states spread across machines.

### 4.1 Single-Pass: Ordinary Least Squares

In ordinary-least-squares (OLS) linear regression the goal is to fit a linear function to a set of points, with the objective of minimizing the sum of squared residuals. Formally, we are given points $(x_1, y_1), \ldots, (x_n, y_n)$, where $x_i \in \mathbb{R}^d$ and $y_i \in \mathbb{R}$, and our goal is to find the vector $\widehat{b}$ that minimizes $\sum_{i=1}^n (y_i - \langle \widehat{b}, x_i \rangle)^2$. OLS is one of the most fundamental methods in statistics. Typically, each $y_i$ is assumed to be an (independent) noisy measurement of $\langle b, x_i \rangle$, where $b$ is an unknown but fixed vector and the noise is uncorrelated with mean 0 and unknown but fixed variance. Under these assumptions, $\widehat{b}$ is the best linear unbiased estimate of $b$ (Gauss-Markov). Under additional assumptions (normality, independence), $\widehat{b}$ is also the maximum-likelihood estimate. Letting $X$ denote the matrix whose rows are $x_i^T$, and defining $y := (y_1, \ldots, y_n)^T$, it is well-known that the sum of squared residuals is minimized by $\widehat{b} = (X^T X)^{-1} X^T y$ (for exposition purposes we assume the full-rank case here, though this is not a requirement for MADlib).

It has been observed before that computing $\widehat{b}$ lends itself well to data-parallel implementations in databases [31, 30] and map-reduce [10]. In extensible-database terms, this task can be done with a simple user-defined aggregate. The principal observation is this: $X^T X = \sum_{i=1}^n x_i x_i^T$ and $X^T y = \sum_{i=1}^n x_i y_i$ are just sums of transformations of each data point. Summation is associative, so data parallelism virtually comes for free—we can compute the per-process subsums of the previous expressions locally in each process, and then sum up all subsums during a second-phase aggregation. As a final non-parallelized step, we compute the inverse of $X^T X$ and then multiply with $X^T y$. These final operations are comparatively cheap, since the number of independent variables (and thus the dimensions of $X^T X$ and $X^T y$) is typically "small".

#### 4.1.1 MADlib Implementation

We assume that data points are stored as tuples (`x DOUBLE PRECISION[]`, `y DOUBLE PRECISION`). Linear regression is then implemented as a user-defined aggregate with a transition and final function roughly as in Listings 1 and 2, respectively. (For compactness, we omitted finiteness checks and several output statistics in the example here.) The merge function, which is not shown, just adds all values in the transition states together. Running the code produces the following:

```
psql# SELECT (linregr(y, x)).* FROM data;
-[ RECORD 1 ]+-------------------------------------------
coef         | {1.7307,2.2428}
r2           | 0.9475
std_err      | {0.3258,0.0533}
t_stats      | {5.3127,42.0640}
p_values     | {6.7681e-07,4.4409e-16}
condition_no | 169.5093
```

Note that the `linregr` Python UDF produces a composite record type in the output, which is a feature of PostgreSQL and Greenplum. This would be easy to flatten into a string in a strictly relational implementation.

### 4.2 Multi-Pass: (Binary) Logistic Regression

In (binary) logistic regression, we are given points $(x_1, y_1), \ldots, (x_n, y_n)$, where $x_i \in \mathbb{R}^d$ and $y_i \in \{0, 1\}$, and our goal is to find the vector $\widehat{b}$ that maximizes $\prod_{i=1}^n \sigma((-1)^{y_i+1} \cdot \langle \widehat{b}, x_i \rangle)$. Here, $\sigma(z) = \frac{1}{1+\exp(z)}$ denotes the logistic function. Statistically, this is the maximum-likelihood estimate for an unknown vector $b$ under the assumption

```
1   AnyType
2   linregr_transition::run(AnyType& args) {
3       // Transition state is a class that wraps an array.
4       // We expect a mutable array. If DBMS allows
5       // modifications, copying will be avoided.
6       LinRegrTransitionState<
7           MutableArrayHandle<double> > state = args[0];
8       // Dependent variable is a double-precision float
9       double y = args[1].getAs<double>();
10      // Vector of independent variables wraps an immutable
11      // array (again, no unnecessary copying). This maps
12      // to an Eigen type
13      MappedColumnVector x
14          = args[2].getAs<MappedColumnVector>();
15
16      if (state.numRows == 0) {
17          // The first row determines the number
18          // of independent variables
19          state.initialize(*this, x.size());
20      state.numRows++;
21      state.y_sum += y;
22      state.y_square_sum += y * y;
23      // noalias informs Eigen to multiply in-place
24      state.X_transp_Y.noalias() += x * y;
25      // Since X^T X is symmetric, we only need to
26      // compute a triangular part
27      triangularView<Lower>(state.X_transp_X)
28          += x * trans(x);
29
30      return state;
31  }
```

**Figure 1: Linear regression transition function**

```
1   AnyType
2   linregr_final::run(AnyType& args) {
3       // Immutable array: Array will never be copied
4       LinRegrTransitionState<ArrayHandle<double> > state
5           = args[0];
6
7       // The following is a MADlib class that wraps Eigen's
8       // solver for self-adjoint matrices.
9       SymmetricPositiveDefiniteEigenDecomposition<Matrix>
10          decomposition(state.X_transp_X, EigenvaluesOnly,
11              ComputePseudoInverse);
12
13      Matrix inverse_of_X_transp_X
14          = decomposition.pseudoInverse();
15      // Let backend allocate array for coefficients so to
16      // avoid copying on return (if supported by DBMS).
17      MutableMappedColumnVector coef(
18          allocateArray<double>(state.widthOfX));
19      coef.noalias() = inverse_of_X_transp_X
20          * state.X_transp_Y;
21
22      // Return a composite value.
23      AnyType tuple;
24      tuple << coef << decomposition.conditionNo();
25      return tuple;
26  }
```

**Figure 2: Linear regression final function**

that each $y_i$ is a random variate with $\Pr[y_i = 1 \mid x_i] = \sigma(\langle b, x_i \rangle)$ and that all observations are independent.

It is well-known that, in general, no closed-formula expression for $\widehat{b}$ exists. Instead, $\widehat{b}$ can be computed as the solution of a convex program via standard iterative methods. Arguably, the most common method is to maximize the logarithm of the likelihood using Newton's method. In the case of logistic regression this reduces to *iteratively reweighted least squares* with iteration rule $\widehat{\beta}_{m+1} = (X^T D_m X)^{-1} X^T D_m z_m$. Here, the diagonal matrix $D_m$ and the vector $z_m$ are transformations of $X$ and $\widehat{\beta}_m$.

### 4.2.1 MADlib Implementation

Each individual iteration can be implemented via a user-defined aggregate using linear regression as a blueprint. However, the handling of iterations and checking for convergence require a further outer loop. We therefore implement a driver UDF in Python. The control flow follows the high-level outline from Section 3.1.2 and is illustrated as an activity diagram in Figure 3. Here, the shaded shapes are executions of generated SQL, where `current_iteration` is a template parameter that is substituted with the corresponding Python variable.

Specifically, the UDF first creates a temporary table for storing the inter-iteration states. Then, the Python code iteratively calls the UDA for updating the iteration state, each time adding a new row to the temporary table. Once the convergence criterion has been reached, the state is converted into the return value. The important point to note is that there is no data movement between the driver function and the database engine—all heavy lifting is done within the database engine.

Unfortunately, implementing logistic regression using a driver function leads to a different interface than the one we provided for linear regression:

```
SELECT * FROM logregr('y', 'x', 'data');
```

A problem with this implementation is that the `logregr` UDF is not an aggregate function and cannot be used in grouping constructs. To perform multiple logistic regressions at once, one needs to use a join construct instead. We intend to address this non-uniformity in interface in a future version of MADlib. We highlight the issue here in part to point out that SQL can be a somewhat "over-rich" language. In many cases there are multiple equivalent patterns for constructing simple interfaces, but no well-accepted, uniform design patterns for the kind of algorithmic expressions we tend to implement in MADlib. We are refining these design patterns as we evolve the library.

## 4.3 Large-State Iteration: k-Means

In $k$-means clustering, we are given $n$ points $x_1, \ldots, x_n \in \mathbb{R}^d$, and our goal is to position $k$ centroids $c_1, \ldots, c_k \in \mathbb{R}^d$ so that the sum of squared distances between each point and its closest centroid is minimized. Formally, we wish to minimize $\sum_{i=1}^{n} \min_{j=1}^{k} \|x_i - c_j\|^2$. Solving this problem exactly is usually prohibitively expensive (for theoretical hardness results see, e.g., [1, 23]). However, the local-search heuristic proposed by Lloyd [21] performs reasonably well both in theory and in practice [5, 4]. At a high level, it works as follows:

1. Seeding phase: Find initial positions for $k$ centroids $c_1, \ldots, c_k$.
2. Assign each point $x_1, \ldots, x_n$ to its closest centroid.
3. Reposition each centroid to the barycenter (mean) of all points assigned to it.
4. If no (or only very few) points got reassigned, stop. Otherwise, goto (2).

### 4.3.1 MADlib implementation

$k$-means has a natural implementation in SQL [29]. Based on the assumption that we can always comfortably store $k$ centroids in main memory, we can implement $k$-means similarly to logistic regression: Using a driver function that iteratively calls a user-defined aggregate. In the following, we take a closer look at this implementation. It is important to make a clear distinction between the inter-iteration state (the output of the UDA's final function) and intra-iteration state (as maintained by the UDA's transition and merge functions). During aggregation, the transition state contains both inter- and

1705

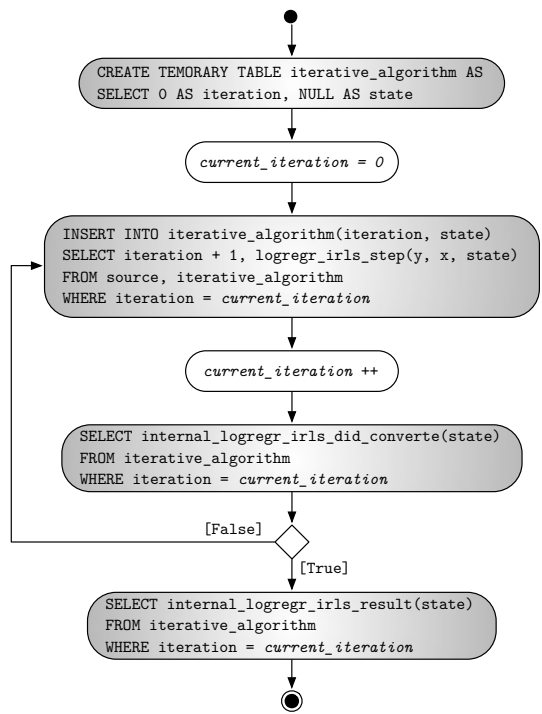

**Figure 3: Sequence diagram for logistic regression**

intra-iteration state, but only modifies the intra-iteration state. We only store $k$ centroids in both the inter- and intra-iteration states, and consider the assignments of points to centroids as implicitly given.

In the transition function, we first compute the centroid that the current point was closest to at the beginning of the iteration using the inter-iteration state. We then update the barycenter of this centroid in the intra-iteration state. Only as the final step of the aggregate, the intra-iteration state becomes the new inter-iteration state.

Unfortunately, in order to check the convergence criterion that no or only few points got reassigned, we have to do two closest-centroid computations per point and iteration: First, we need to compute the closest centroid in the previous iteration and then the closest one in the current iteration. If we stored the closest points explicitly, we could avoid half of the closest-centroid calculations.

We can store points in a table called `points` that has a `coords` attribute containing the points' coordinates and has a second attribute for the current `centroid_id` for the point. The iteration state stores the centroids' positions in a matrix called `centroids`. MADlib provides a UDF `closest_column(a,b)` that determines the column in a matrix `a` that is closest to vector `b`. Thus, we can make the point-to-centroid assignments explicit using the following SQL:

```
UPDATE points
  SET centroid_id = closest_column(centroids, coords)
```

Ideally, we would like to perform the point reassignment and the repositioning with a single pass over the data. Unfortunately, this cannot be expressed in standard SQL.[1]

---
[1]While PostgreSQL and Greenplum provide an optional RETURNING clause for UPDATE commands, this returns only one row for each row affected by the UPDATE, and aggregates cannot be used within the RETURNING clause. Moreover, an UPDATE ... RETURNING cannot be used as a subquery.

Therefore, while we can reduce the number of closest-centroid calculations by one half, PostgreSQL processes queries one-by-one (and does not perform cross-statement optimization), so it will need to make two passes over the data per one $k$-means iteration. In general, the performance benefit of explicitly storing points depends on the DBMS, the data, and the operating environment.

The pattern of updating temporary state is made a bit more awkward in PostgreSQL due to its legacy of versioned storage. PostgreSQL performs an update by first inserting a new row and then marking the old row as invisible [39, Section 23.1.2]. As a result, for updates that touch many rows it is typically faster to copy the updated data into a new table (i.e., CREATE TABLE AS SELECT and DROP TABLE) rather than issue an UPDATE. These kinds of DBMS-specific performance tricks may merit encapsulation in an abstraction layer for SQL portability.

## 4.4 Infrastructure Performance Trends

In its current beta version, MADlib has been tuned a fair bit over PostgreSQL and Greenplum, though much remains to be done. Here we report on some results for a basic scenario that exercises our core functionality, including the C++ abstraction layer, our ability to call out to linear-algebra packages, and parallel-speedup validation. We defer macro-benchmarking of MADlib's current methods to future work, that will focus on specific algorithm implementations.

The basic building block of MADlib is a user-defined aggregate, typically one that calls out to a linear-algebra library. In order to evaluate the scalability of this construct, we ran linear regression over Greenplum's parallel DBMS on various data sizes, using a 24-core test cluster we had available, which was outfitted with 144 GB of RAM over 51 TB of raw storage.[2] This is obviously a relatively modest-sized cluster by today's standards, but it is sufficient to illuminate (a) our efforts at minimizing performance overheads, and (b) our ability to achieve appropriate parallel speedup.

For running linear regression as outlined in Section 4.1, we expect runtime $O(k^3 + (n \cdot k^2)/p)$ where $k$ is the number of independent variables, $n$ is the number of observations, and $p$ is the number of query processes. The $k^3$ time is needed for the matrix inversion, and the $k^2$ is needed for computing each outer product $x_i x_i^T$ and adding it to the running sum. It turns out that our runtime measurements fit these expectations quite well, and the constant factors are relatively small. See Figures 4 and 5. In particular we note:

- The overhead for a single query is very low and only a fraction of a second. This also implies that we lose little in implementing iterative algorithms using driver functions that run multiple SQL queries.

- Given the previous points, the Greenplum database achieves perfect linear speedup in the example shown.

In the example, all data was essentially in the buffer cache, and disk I/O was not a limiting factor. This is typical in large parallel installations, since an analyst often runs a machine-learning algorithm several times with different parameters. Given large amounts of main memory on each node in a well-provisioned cluster, much of the data should remain in the cache. Finally, the computational cost per row grows at least quadratically, and thus will easily surpass I/O cost for complex models. As we compared various cluster sizes, numbers of independent variables and the respective execution times for previous versions of MADlib, the lesson we learned is that even though we anticipate non-trivial overhead by the DBMS, careful performance tuning—e.g., by making use of instruction-accurate profiling using Valgrind [27]—still makes significant differences:

- Version 0.1alpha is an implementation in C that computes the outer-vector products $x_i x_i^T$ as a simple nested loop.

- Version 0.2.1beta introduced an implementation in C++ that used the Armadillo [37] linear-algebra library as a frontend for LAPACK/BLAS. It turns out that this version was much slower for two reasons: The BLAS library used was the default one shipped with CentOS 5, which is built from the untuned reference BLAS. Profiling and examining the critial code paths revealed that computing $y^T y$ for a row vector $y$ is about three to four times slower than computing $x x^T$ for a column vector $x$ of the same dimension (and the MADlib implementation unfortunately used to do the former). Interestingly, the same holds for Apple's Accelerate framework on Mac OS X, which Apple promises to be a tuned library. The second reason for the speed disadvantage is runtime overhead in the first incarnation of the C++ abstraction layer (mostly due to locking and calls into the DBMS backend).

- Version 0.3 has an updated linear-regression implementation that relies on the Eigen C++ linear-algebra library and takes advantage of the fact that the matrix $X^T X$ is symmetric positive definite. Runtime overhead has been reduced, but some calls into the database backend still need better caching.

---
[2]Our cluster is made up of four SuperMicro X8DTT-H server modules, each equipped with one six-core Intel Xeon X5670 processor, clocked at 2.93 GHz. While hyperthreading is enabled, we only run a single Greenplum "segment" (query process) per physical core. Each machine has 24 GB of RAM, an LSI MegaRAID 2108 "Raid On a Chip" controller with six attached 360 GB solid-state drives, and a Brocade 1020 converged network adapter. The operating system is Red Hat Enterprise Linux Server release 5.5 (Tikanga). On that we are running Greenplum Database 4.2.0, compiled with gcc 4.4.2.

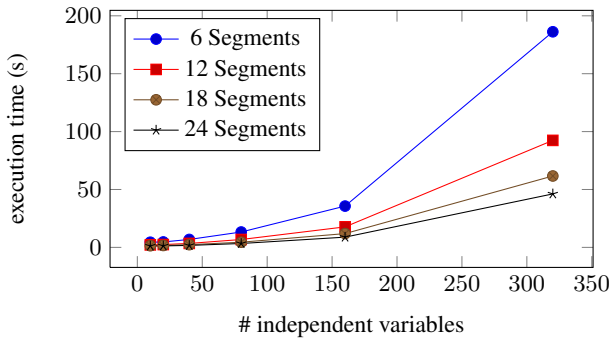

Figure 5: Linear regression execution times using MADlib v0.3 on Greenplum Database 4.2.0, 10 million rows

Other noteworthy results during our performance studies included that there are no measurable performance differences between PostgreSQL 9.1.1 (both in single and multi-user mode) and GP 4.1 in running the aggregate function on a single core. Moreover, while testing linear/logistic-regression execution times, single-core performance of even laptop CPUs (like the Core i5 540M) did not differ much from today's server CPUs (like the Xeon family). Typically the differences were even less than what the difference in clock speeds might have suggested, perhaps due to compiler and/or architecture issues that we have yet to unpack.

## 5. UNIVERSITY RESEARCH AND MADLIB

An initial goal of the MADlib project was to engage closely with academic researchers, and provide a platform and distribution channel for their work. To that end, we report on two collaborations over the past years. We first discuss a more recent collaboration with researchers at the University of Wisconsin. We then report on a collaboration with researchers at Florida and Berkeley, which co-evolved with MADlib and explored similar issues (as part of the BayesStore project [44, 43]) but was actually integrated with the MADlib infrastructure only recently.

### 5.1 Wisconsin Contributions: Convex Optimization

The MADlib framework goes a long way toward making in-database analytic tools easier to deploy inside an RDBMS. Nevertheless, to implement an algorithm within MADlib, a developer must undertake several steps: they must specify a model, select the algorithm used to implement that model, optimize the algorithm, test the model, and finally ensure that the resulting algorithm and model are robust. This is a time consuming process that creates code that must be tested and maintained for each data analysis technique.

A focus of the MADlib work at the University of Wisconsin has been to explore techniques to reduce this burden. Our approach is to expose a mathematical abstraction on top of MADlib that allows a developer to specify a smaller amount of code, which we hope will lower the development time to add new techniques in many cases. We discuss the challenge that we faced implementing this abstraction. To demonstrate our ideas, we have implemented all of the models shown in Table 2 within the single abstraction (built within MADlib) that we describe below.

*The Key Abstraction.* An ideal abstraction would allow us to decouple the specification of the model from the algorithm used to solve the specification. Fortunately, there is a beautiful, powerful abstraction called convex optimization that has been developed for the last few decades [36, 8] that allows one to perform this decoupling. More precisely, in convex optimization, we minimize a convex function over a convex set. The archetype convex function is $f(x) = x^2$ and is shown in Figure 6. Like all convex functions, any local minimum of $f$ is a global minimum of $f$. Many different popular statistical models are defined by convex optimization problems, e.g., linear regression, support vector machines, logistic regression, conditional random fields. Not every data analysis problem is convex—notable exceptions include the a priori algorithm and graph mining algorithms—but many are convex. Table 2 lists models that we have implemented in our abstraction over MADlib, all of which are convex. (Note that the previous built-in MADlib examples of linear and logistic regression fall into this category!)

In spite of the expressiveness of convex optimization, even simple algorithms converge at provable rates to the true solution. For intuition, examine the archetypical function $f(x) = x^2$ shown in Figure 6. The graph of this function is like all convex sets: bowl shaped. To minimize the function, we just move to the bottom of the bowl. As a result, even greedy schemes that decrease the function at each step will converge to an optimal solution. One such popular greedy method is called a gradient method. The idea is to find the steepest descent direction. In 1-d, this direction is the opposite direction of the derivative; in higher dimensions, it is called the gradient of $f$. Using the gradient, we iteratively move toward a solution. This process can be described by the following pseudocode:

$$x \leftarrow x - \alpha \cdot G(x)$$

where $G(x)$ is the gradient of $f(x)$ and $\alpha$ is a positive number called the *stepsize* that goes to zero with more iterations. For example, it suffices to set $\alpha = 1/k$ where $k$ is the number of iterations. In the $f(x) = x^2$ example, the gradient is the derivative, $G(x) = 2x$. Since $x = 0$ is the minimum value, we have that for $x > 0$, $G(x) < 0$ while for $x < 0$ we have that $G(x) > 0$. For a convex function, the gradient always tells us which direction to go to find a minimum value, and the process described above is guaranteed to converge at a known rate. One can provide a provable rate of convergence to the minimum value, which is in sharp contrast to a typical greedy search. In our prototype implementation in MADlib, we picked up one such simple greedy algorithm, called stochastic (or sometimes, "incremental") gradient descent (SGD) [35, 6], that goes back to the 1960s. SGD is an approximation of gradient methods that is useful when the convex function we are considering, $f(x)$, has the form:

$$f(x) = \sum_{i=1}^{N} f_i(x)$$

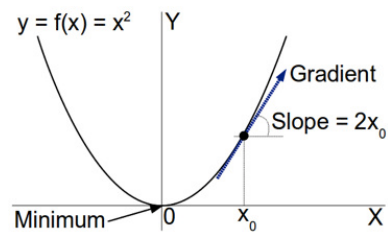

Figure 6: The archetypical convex function $f(x) = x^2$.

| # segments | # variables | # rows (million) | v0.3 (s) | v0.2.1beta (s) | v0.1alpha (s) |
|---|---|---|---|---|---|
| 6 | 10 | 10 | 4.447 | 9.501 | 1.337 |
| 6 | 20 | 10 | 4.688 | 11.60 | 1.874 |
| 6 | 40 | 10 | 6.843 | 17.96 | 3.828 |
| 6 | 80 | 10 | 13.28 | 52.94 | 12.98 |
| 6 | 160 | 10 | 35.66 | 181.4 | 51.20 |
| 6 | 320 | 10 | 186.2 | 683.8 | 333.4 |
| 12 | 10 | 10 | 2.115 | 4.756 | 0.9600 |
| 12 | 20 | 10 | 2.432 | 5.760 | 1.212 |
| 12 | 40 | 10 | 3.420 | 9.010 | 2.046 |
| 12 | 80 | 10 | 6.797 | 26.48 | 6.469 |
| 12 | 160 | 10 | 17.71 | 90.95 | 25.67 |
| 12 | 320 | 10 | 92.41 | 341.5 | 166.6 |
| 18 | 10 | 10 | 1.418 | 3.206 | 0.6197 |
| 18 | 20 | 10 | 1.648 | 3.805 | 1.003 |
| 18 | 40 | 10 | 2.335 | 5.994 | 1.183 |
| 18 | 80 | 10 | 4.461 | 17.73 | 4.314 |
| 18 | 160 | 10 | 11.90 | 60.58 | 17.14 |
| 18 | 320 | 10 | 61.66 | 227.7 | 111.4 |
| 24 | 10 | 10 | 1.197 | 2.383 | 0.3904 |
| 24 | 20 | 10 | 1.276 | 2.869 | 0.4769 |
| 24 | 40 | 10 | 1.698 | 4.475 | 1.151 |
| 24 | 80 | 10 | 3.363 | 13.35 | 3.263 |
| 24 | 160 | 10 | 8.840 | 45.48 | 13.10 |
| 24 | 320 | 10 | 46.18 | 171.7 | 84.59 |

**Figure 4: Linear regression execution times**

If each of the $f_i$ is convex, then so is $f$ [8, pg. 38]. Notice that all problems in Table 2 are of this form: intuitively each of these models is finding some model (i.e., a vector $w$) that is scored on many different training examples. SGD leverages the above form to construct a rough estimate of the gradient of $f$ using the gradient of a single term: for example, the estimate if we select $i$ is the gradient of $f_i$ (that we denote $G_i(x)$). The resulting algorithm is then described as:

$$x \leftarrow x - \alpha N \cdot G_i(x) \qquad (1)$$

This approximation is guaranteed to converge to an optimal solution [26].

*Using the MADlib framework.* In our setting, each tuple in the input table for an analysis task encodes a single $f_i$. We use the *micro-programming* interfaces of Sections 3.2 and 3.3 to perform the mapping from the tuples to the vector representation that is used in Eq. 1. Then, we observe Eq. 1 is simply an expression over each tuple (to compute $G_i(x)$) which is then averaged together. Instead of averaging a single number, we average a vector of numbers. Here, we use the *macro-programming* provided by MADlib to handle all data access, spills to disk, parallelized scans, etc. Finally, the macro programming layer helps us test for convergence (which is implemented with either a python combination or C driver.) Using this approach, we were able to add in implementations of all the models in Table 2 in a matter of days.

In an upcoming paper we report initial experiments showing that our SGD based approach achieves higher performance than prior data mining tools for some datasets [13].

### 5.2 Florida/Berkeley Contributions: Statistical Text Analytics

The focus of the MADlib work at Florida and Berkeley has been to integrate statistical text analytics into a DBMS. In many domains, structured data and unstructured text are both important assets for

1708

| Application | Objective |
|---|---|
| Least Squares | $\sum_{(u,y)\in\Omega}(x^T u - y)^2$ |
| Lasso [40] | $\sum_{(u,y)\in\Omega}(x^T u - y)^2 + \mu\|x\|_1$ |
| Logisitic Regression | $\sum_{(u,y)\in\Omega}\log(1 + \exp(-yx^T u))$ |
| Classification (SVM) | $\sum_{(u,y)\in\Omega}(1 - yx^T u)_+$ |
| Recommendation | $\sum_{(i,j)\in\Omega}(L_i^T R_j - M_{ij})^2 + \mu\|L, R\|_F^2$ |
| Labeling (CRF) [42] | $\sum_k \left[\sum_j x_j F_j(y_k, z_k) - \log Z(z_k)\right]$ |

**Table 2: Models currently Implemented in MADlib using the SGD-based approach.**

| Statistical Methods | POS | NER | ER |
|---|---|---|---|
| Text Feature Extraction | ✓ | ✓ | ✓ |
| Viterbi Inference | ✓ | ✓ | |
| MCMC Inference | | ✓ | ✓ |
| Approximate String Matching | | | ✓ |

**Table 3: Statistical text analysis methods**

data analysis. For example, electronic medical record systems use transactional databases to store both structured data such as lab results and monitor readings, and unstructured data such as notes from physicians. With our goal to push analysis close to the data, it is important for MADlib to support basic statistical methods for data scientists to implement text analysis tasks.

Basic text analysis tasks include part-of-speech (POS) tagging, named entity extraction (NER), and entity resolution (ER) [12, 18]. Different statistical models and algorithms are implemented for each of these tasks with different runtime-accuracy tradeoffs. For example, an entity resolution task could be to find all mentions in a text corpus that refer to a real-world entity *X*. Such a task can be done efficiently by approximate string matching [25] techniques to find all mentions in text that approximately match the name of entity *X*. However, such a method is not as accurate as the state-of-the-art collective entity resolution algorithms based on statistical models, such as Conditional Random Fields (CRFs) [19].

*Pushing Statistical Text Analysis into MADlib.* Based on the MADlib framework, our group set out to implement statistical methods in SQL to support various text analysis tasks. We use CRFs as the basic statistical model to perform more advanced text analysis. Similar to Hidden Markov Models (HMM) cite, CRFs are a leading probabilistic model for solving many text analysis tasks, including POS, NER and ER [19]. To support sophisticated text analysis, we implement four key methods: text feature extraction, most-likely inference over a CRF (Viterbi), MCMC inference, and approximate string matching (Table 3).

**Text Feature Extraction:** Text feature extraction is a step in most statistical text analysis methods, and it can be an expensive operation. To achieve high quality, CRF methods often assign hundreds of features to each token in the document. Examples of such features include: (1) dictionary features: *does this token exist in a provided dictionary?* (2) regex features: *does this token match a provided regular expression?* (3) edge features: *is the label of a token correlated with the label of a previous token?* (4) word features: *does this the token appear in the training data?* and (5) position features: *is this token the first or last in the token sequence?* The right combination of features depends on the application, and so leveraging the micro-programming interface, we implemented a common set of text feature extractors.

**Approximate String Matching:** A recurring primitive operation in text processing applications is the ability to match strings approximately. The technique we use is based on qgrams [16]. We used the trigram module in PostgreSQL to create and index 3-grams over text. Given a string "Tim Tebow" we can create a 3-gram by using a sliding window of 3 characters over this text string. Using the 3-gram index, we created an approximate matching UDF that takes in a query string and returns all documents in the corpus that contain at least one approximate match.

Once we have the features, the next step is to perform inference on the model. We also implemented two types of statistical inference within the database: Viterbi (when we only want the most likely answer from the model) and MCMC (when we want the probabilities or confidence of an answer as well).

**Viterbi Inference:** The *Viterbi* dynamic programming algorithm is a popular algorithm to find the top-k most likely labelings of a document for (linear chain) CRF models [14].

Like any dynamic programming algorithm, the Viterbi algorithm is recursive. We experimented with two different implementations of macro-coordination over time. First, we chose to implement it using a combination of recursive SQL and window aggregate functions. We discussed this implementation at some length in earlier work [44]. Our initial recursive SQL implementation only runs over PostgreSQL versions 8.4 and later; it does not run in Greenplum. Second, we chose to implement a Python UDF that uses iterations to drive the recursion in Viterbi. This iterative implementation runs over both PostgreSQL and Greenplum. In Greenplum, Viterbi can be run in parallel over different subsets of the document on a multi-core machine.

**MCMC Inference:** Markov chain Monte Carlo (MCMC) methods are classical sampling algorithms that can be used to estimate probability distributions. We implemented two MCMC methods in MADlib: Gibbs sampling and Metropolis-Hastings (MCMC-MH).

The MCMC algorithms involve iterative procedures where the current iteration depends on previous iterations. We used SQL window aggregates for macro-coordination in this case, to carry "state" across iterations to perform the Markov-chain process. This window function based implementation runs over PostgreSQL version 8.4 and later. We discussed this implementation at some length in recent work [43]. We are currently working on integrating MCMC algorithms into Greenplum DBMS. We also plan to implement MCMC using Python UDF macro-coordination analogous to Section 4.3, and compare the performance between the two implementations.

*Using the MADlib Framework.* Because this work predated the release of MADlib, it diverged from the macro-coordination patterns of Section 3.1, and took advantage of PostgreSQL features that are less portable. These details require refactoring to fit into the current MADlib design style, which we are currently working on. In addition to the work reported above, there are a host of other features of both PostgreSQL and MADlib that are valuable for text analytics. Extension libraries for PostgreSQL and Greenplum provides text processing features such as inverted indexes, trigram indexes for approximate string matching, and array data types for model parameters. Existing modules in MADlib, such as Naive Bayes and Sparse/Dense Matrix manipulations, are building blocks to implement statistical text analysis methods. Leveraging this diverse set of tools and techniques that are already within the database allowed us to build a sophisticated text analysis engine that has comparable raw performance to off-the-shelf implementations, but runs natively in a DBMS close to the data [44, 43].

## 6. RELATED WORK

The space of approaches to marry analytics with data is large and seems to be growing rapidly. At a high-level there are two approaches, (1) bring a statistical language to a data processing

1709

substrate, or (2) provide a framework to express statistical techniques on top of a data processing substrate. Specifically, we split the approaches to achieve this marriage into two groups: (1) *top-down approaches* begin with a high-level statistical programming language, e.g., R or Matlab. The technical goal is to build a parallel data processing infrastructure that is able to support running this high-level language in parallel. Examples of approach include System ML from IBM [15], Revolution Analytics [33], and SNOW [41]. (2) We call the second group *framework-based approaches* whose goal is to provide a set of building blocks (individual machine-learning algorithms) along with library support for micro- and macro-programming to write the algorithms. Recent examples are Mahout [3], Graphlab [22], SciDB [38] and MADlib.

*Top-Down, Language-based Approaches.* Within the approaches of type (1), we again divide the space into imperative and declarative approaches. In the imperative approach, an analyst takes the responsibility of expressing how to parallelize the data access, e.g., SNOW and Parallel R packages provide an MPI-interface along with calls for data partitioning within the framework of R. In contrast in a declarative approach, the analyst declares their analysis problem and it is the responsibility of the system to achieve this goal (in analogy with a standard RDBMS). Examples of the declarative approach include SystemML [15], which is an effort from IBM to provide an R-like language to specify machine learning algorithms. In SystemML, these high-level tasks are then compiled down to a Hadoop-based infrastructure. The essence of SystemML is its compilation techniques: they view R as a declarative language, and the goal of their system is to compile this language into an analog of the relational algebra. Similar approaches are taken in Revolution Analytics [33] and Oracle's new parallel R offering [28] in that these approaches attempt to automatically parallelize code written in R.

*Framework-based Approaches.* Framework-based approaches attempt to provide the analyst with low-level primitives and co-ordination primitives built on top of a data processing substrate. Typically, framework-based approaches offer a template whose goal is to automate the common aspects of deploying an analytic task. Framework-based approaches differ in what their data-processing substrates offer. For example, MADlib is in in this category as it provides a library of functions over an RDBMS. The macro- and micro-programming described earlier are examples of design templates. The goal of Apache Mahout [3] is to provide an open-source machine learning library over Apache Hadoop. Currently, Mahout provides a library of machine learning algorithms and a template to extend this library. At a lower level, the DryadLINQ large vector library provides the necessary data types to build analysis tasks (e.g., vectors) that are commonly used by machine learning techniques. SciDB advocates a completely rewritten DBMS engine for numerical computation, arguing that RDBMSs have "the wrong data model, the wrong operators, and are missing required capabilities" [38]. MADlib may be seen as a partial refutation of this claim. Indeed, the assertion in [38] that "RDBMSs, such as GreenPlum [sic]...must convert a table to an array inside user-defined functions" is incorrect: the MADlib C++ library hides representation from the UDF developer and only pays for copies when modifying immutable structures. GraphLab is a framework to simplify the design of programming parallel machine learning tasks. The core computational abstraction is a graph of processing nodes that allows one to define asynchronous updates. Graphlab initially focused on providing a framework for easy access to multicore parallelism [22]. However, the core computational abstraction is general and is now also deployed in a cluster setting. Another popular toolkit in this space is Vowpal Wabbit [20] that is extensively used in the academic machine learning community and provides high performance.

Two recent efforts to provide Scala-based domain-specific languages (DSLs) and parallel execution frameworks blur distinctions between frameworks and language-based approaches. Spark [46] is a Scala DSL targeted at Machine Learning, providing access to the fault-tolerant, main-memory *resilient distributed datasets*: which are read-only collections of data that can be partitioned across a cluster. ScalOps [45] provides a Scala DSL for Machine Learning that is translated to Datalog, which is then optimized to run in parallel on the Hyracks infrastructure [7]. Given its roots in Datalog and parallel relational algebra, ScalOps bears more similarity to MADlib than any of the other toolkits mentioned here. It would be interesting to try and compile its DSL to the MADlib runtime.

*Other Data Processing Systems.* There are a host of data processing techniques that can be used to solve elements of the underlying data analysis problem. For example, Pregel [24] from Google is designed for data analysis over graphs. In Pregel, the abstraction is to write ones code using a graph-based abstraction: each function can be viewed as a node that sends and receives messages to its neighbors in the graph. Pregel distributes the computation and provides fault tolerance.

## 7. CONCLUSION AND DISCUSSION

Scalable analytics are a clear priority for the research and industrial communities. MADlib was designed to fill a vacuum for scalable analytics in SQL DBMSs, and connect database research to market needs. In our experience, a parallel DBMS provides a very efficient and flexible dataflow substrate for implementing statistical and analytic methods at scale. Standardized support for SQL extensions across DBMSs could be better—robust and portable support for recursion, window aggregates and linear-algebra packages would simplify certain tasks. Since this is unlikely to occur across vendors in the short term, we believe that MADlib can evolve reasonable workarounds at the library and "design pattern" level.

The popular alternative to a DBMS infrastructure today is Hadoop MapReduce, which provides much lower-level programming APIs than SQL. We have not yet undertaken performance comparisons with Hadoop-based analytics projects like Mahout. Performance comparisons between MADlib and Mahout today would likely boil down to (a) variations in algorithm implementations, and (b) well-known (and likely temporary) tradeoffs between the current states of DBMSs and Hadoop: C vs. Java, pipelining vs. checkpointing, etc. [32] None of these variations and tradeoffs seem endemic, and they may well converge over time. Moreover, from a marketplace perspective, such comparisons are not urgent: many users deploy both platforms and desire analytics libraries for each. So a reasonable strategy for the community is to foster analytics work in both SQL and Hadoop environments, and explore new architectures (GraphLab, SciDB, ScalOps, etc.) at the same time.

MADlib has room for growth in multiple dimensions. The library infrastructure itself is still in beta, and has room to mature. There is room for enhancements in its core treatment of mathematical kernels (e.g., linear algebra over both sparse and dense matrices) especially in out-of-core settings. And of course there will always be an appetite for additional statistical models and algorithmic methods, both textbook techniques and cutting-edge research. Finally, there is the challenge of porting MADlib to DBMSs other than PostgreSQL and Greenplum. As discussed in Section 3, MADlib's macro-coordination logic is written in largely standard Python and SQL, but its finer grained "micro-programming" layer exploits proprietary DBMS extension interfaces. Porting MADlib across DBMSs is a mechanical but non-trivial software development effort. At the macro level, porting will involve the package infrastructure (e.g. a cross-platform installer) and software engineering framework (e.g. testing scripts for additional database engines). At the micro-programming logic level, inner loops of various methods will need to be revisited (particularly user-defined functions). Finally, since Greenplum retains most of the extension interfaces exposed by PostgreSQL, the current MADlib portability interfaces (e.g., the C++ abstraction of Section 3.3) will likely require revisions when porting to a system without PostgreSQL roots.

Compared to most highly scalable analytics packages today, MADlib v0.3 provides a relatively large number of widely-used statistical and analytic methods. It is still in its early stages of development, but is already in use both at research universities and at customer sites. As the current software matures, we hope to foster more partnerships with academic institutions, database vendors, and customers. In doing so, we plan to pursue additional analytic methods prioritized by both research "push" and customer "pull". We also look forward to ports across DBMSs.

## 8. ACKNOWLEDGMENTS

In addition to the authors, the following people have contributed to MADlib and its development: Jianwang Ao, Joel Cardoza, Brian Dolan, Hlya Emir-Farinas, Christan Grant, Hitoshi Harada, Steven Hillion, Luke Lonergan, Gavin Sherry, Noelle Sio, Kelvin So, Gavin Yang, Ren Yi, Jin Yu, Kazi Zaman, Huanming Zhang.